\documentclass[useAMS,usenatbib]{mn2e}

\usepackage{times}

\usepackage{graphicx}
\usepackage{times}
\newcommand{\cthead}[1]{\multicolumn{1}{c}{#1}}
\newcommand{\kss}{km~s$^{-1}$ }
\newcommand{\ks}{km~s$^{-1}$}
\title[High-velocity  class~I methanol maser]{High-velocity feature of the class~I methanol maser in G309.38$-$0.13}
\author[M. A. Voronkov et al.]{M. A. Voronkov$^{1,2}$\thanks{E-mail:
Maxim.Voronkov@csiro.au}, J. L. Caswell$^1$, T. R. Britton$^{3,1}$, J. A. Green$^{1}$, 
A. M. Sobolev$^4$, \newauthor S. P. Ellingsen$^{5}$\\
$^{1}$Australia Telescope National Facility, CSIRO Astronomy and Space Science, PO Box 76, Epping,
NSW 1710, Australia\\
$^{2}$Astro Space Centre, Profsouznaya st. 84/32, 117997 Moscow, Russia\\
$^{3}$Macquarie University, Department of Physics and Engineering, NSW 2109, Australia\\
$^{4}$Ural State University, Lenin ave. 51, 620083 Ekaterinburg, Russia\\
$^{5}$School of Mathematics and Physics, University of Tasmania, GPO Box
252-37, Hobart, Tasmania 7000, Australia\\}
\begin{document}

\date{}

\pagerange{\pageref{firstpage}--\pageref{lastpage}} \pubyear{2010}

\maketitle

\label{firstpage}

\begin{abstract}
The Australia Telescope Compact Array (ATCA) has been used to map class~I
methanol masers at 36 and 44~GHz in G309.38$-$0.13. Maser spots are found at
nine locations in an area of 
50\arcsec$\times$30\arcsec, with both transitions  reliably detected at only two locations.
The brightest spot is associated with shocked gas
traced by 4.5-$\mu$m emission. The data allowed us to make a serendipitous
discovery of a high-velocity 36-GHz spectral feature, which is blue-shifted by 
about 30~\kss from the peak velocity at this frequency, but spatially located 
close to (within a few arcseconds of) the brightest maser spot.
We interpret this as indicating an outflow parallel to the 
line of sight. Such a high velocity spread of maser features, which has not 
been previously reported in the class~I methanol masers associated with a single 
molecular cloud, suggests that the outflow most likely interacts with a moving parcel of gas.
\end{abstract}


\section{Introduction}
Methanol masers are commonly found in massive star-forming regions,
with more than twenty different centimetre and millimetre wavelength
masing transitions discovered to date \citep[e.g.,][]{mul04}. All methanol
maser transitions do not share the same behaviour. Empirically, they form 
two classes \citep{bat87}. Class~I methanol masers
(e.g. at  36, 44, 84, and 95~GHz) usually occur in multiple locations
across the star-forming region scattered around an area up to a parsec in 
extent \citep[e.g.,][]{kur04,vor06,cyg09}. In contrast, class~II  methanol masers 
(e.g. at 6.7, 12 and 107~GHz) reside in the close vicinity of exciting young stellar objects (YSOs)
and are typically found as a single cluster of emission at arcsecond resolution 
\citep[e.g.,][]{phi98}. Theoretical calculations are able to explain 
this empirical classification and strongly suggest that the pumping process of 
class~I masers is dominated by collisions with molecular hydrogen, in contrast to 
class~II masers which are pumped by radiative excitation 
\citep[e.g.][and references therein]{vor05}.

The morphology of class~I methanol masers has recently become the focus of high angular 
resolution studies aimed at
searching for associations with other phenomena commonly observed
in regions of high-mass star-formation \citep[e.g.,][]{cyg09}. The common consensus is that the 
majority of class~I masers trace interface regions between outflows and molecular gas,
although direct observational evidence of this has been obtained for a limited number of 
sources only \citep[e.g.,][]{pla90,kur04,vor06}. The alternative scenarios involving
cloud-cloud collisions \citep[e.g.,][]{sob92,sal02,sjo10} as well as the interaction of 
expanding H{\sc ii} regions with the ambient molecular environment  \citep{vor10} may also 
be realised in some sources.
The common point of all these scenarios is the presence of shocked gas, where the physical 
conditions are favouring class~I methanol masers (see the discussion in 
section~\ref{hv_discussion} below). Apart from the outflow associations cited above
(based on the 2.12-$\mu$m H$_2$ emission, which is a well known shock tracer),
\citet{cyg09} reported
association of some class~I maser spots  with the  extended features 
showing a prominent excess of  the 4.5-$\mu$m emission in the images obtained
with the Spitzer Space Telescope's Infrared Array Camera (IRAC), also known as
extended green objects \citep[EGOs;][]{cyg09} or ``green fuzzies'' \citep{cha09}. The excess of  
the 4.5-$\mu$m (IRAC band 2) emission could be a result of shock excitation of
molecular hydrogen and carbon monoxide in protostellar outflows 
\citep{cyg09,cha09,deb10}.  
It is worth noting that \citet{che09} demonstrated statistically 
the presence of an EGO in the vicinity of a large fraction of class~I methanol masers at low angular resolution (single dish positions). To increase the number of class~I masers studied at high 
angular resolution and to compare the morphologies observed in different maser transitions 
we carried out an interferometric survey at 36 and 44~GHz of all class~I masers previously reported
in the literature and located south of declination $-$35\degr. In this letter, we present a study of 
G309.38$-$0.13, an especially interesting source from the survey. It represents the level of
morphological and kinematical complexity encountered for the majority of sources observed in
the survey, but it has a distinct high-velocity maser feature at 36-GHz, which was quite 
unexpected \citep[see e.g.,][]{bac90,kal10}.  This maser is found in an area devoid of bright 
radio-continuum emission, but near infrared sources which are presumably embedded stars.
It is located approximately 20\arcmin{ }offset from the prominent H{\sc ii} region Gum~48d
at the heliocentric distance of about 3.5~kpc \citep[for further information on the region see][]{kar09}.

\section{Observations}

\begin{figure}
\includegraphics[width=\linewidth]{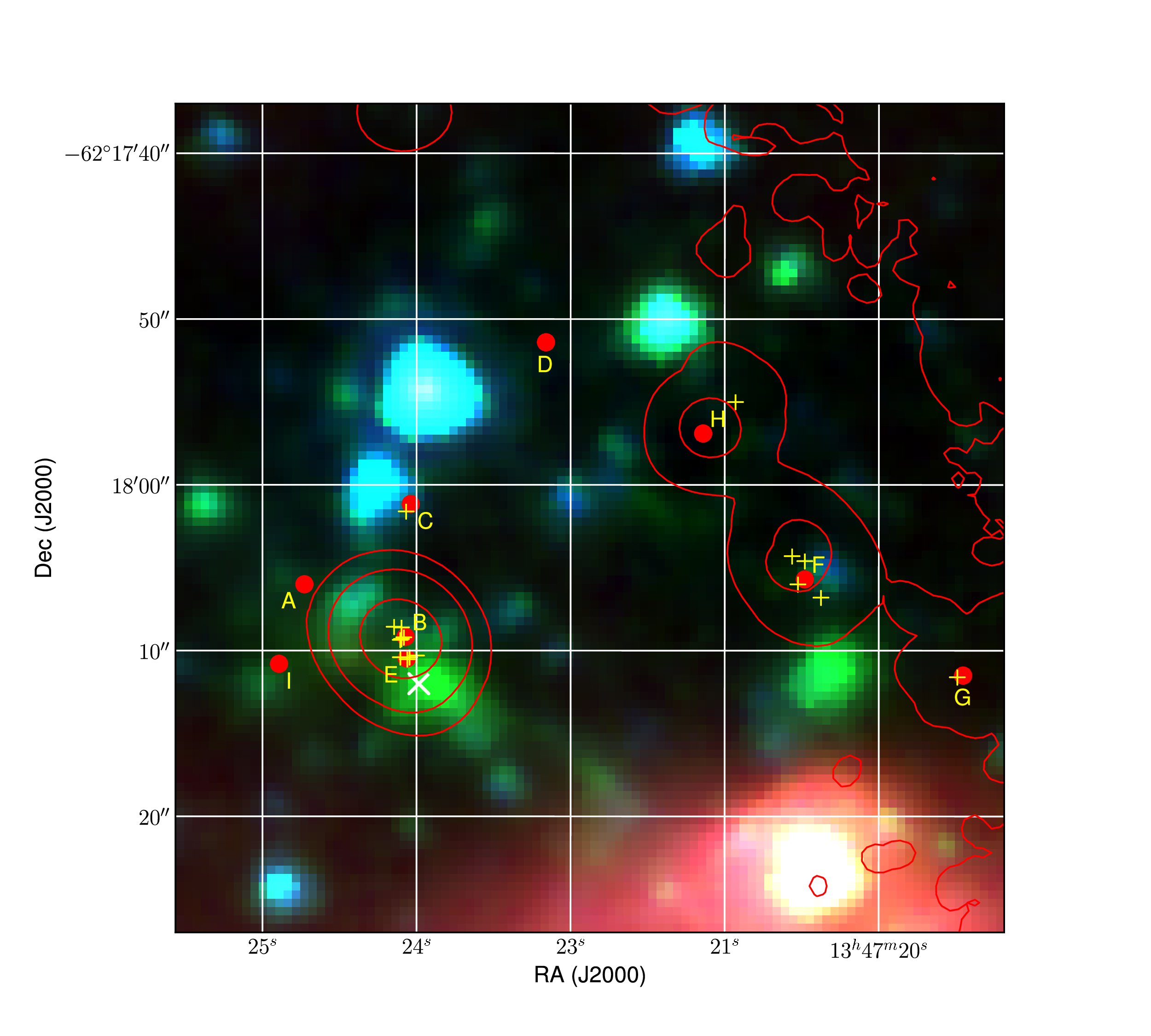}
\caption{Positions of class~I methanol masers (red filled circles and yellow crosses)  overlaid on top of the 3-colour Spitzer image of the G309.38$-$0.13 region. The emission in  8.0, 4.5 and 3.6~$\mu$m IRAC Spitzer bands is shown as red, green and blue, respectively. The filled red circles show the fitted position of the strongest spectral component at each location, while the 
weaker components are shown by yellow crosses (all components are 
given in Table~\protect\ref{fit_results}). The location of a class~II 6.7-GHz methanol
maser \protect\citep{cas09} is shown by the white cross.
The contours  show the maximum of the 36-GHz emission across all spectral
channels. The levels are   5, 15, 50 and 90 per cent of the peak 36-GHz flux density
of 16.1~Jy. Low level artefacts in the top right corner are caused by the primary beam correction.}
\label{glm_mas_overlay}
\end{figure}

Observations were made with the Australia Telescope Compact Array (ATCA) in May 2007 
as part of the interferometric survey of southern class~I methanol masers at 36 and 44~GHz 
(project code C1642). The source was observed using 
the hybrid H214C array configuration in a few  3 minute cuts (four cuts at 36~GHz and 
six at 44~GHz) spread in hour angle for better uv-coverage.  
We did not use, for imaging, the data from the CA06 antenna (located 
about 4.5~km to the west of the other 5 antennas). The remaining antennas provided 
baselines ranging from 82 to 240~m.  
The position of the phase and pointing centre was
$\alpha_{2000}=13^h47^m$26\fs01, $\delta_{2000}=-62$\degr18\arcmin11\farcs66.   
We used reference pointing procedures and determined corrections using  the continuum 
source 1414-59 (which served also as a phase calibrator). From the statistics
of pointing solutions the reference pointing accuracy was estimated
to be 4\farcs8$\pm$2\farcs6. This accuracy affects the accuracy of flux density measurements,
particularly for sources of emission which are offset from the pointing centre.  The
positional accuracy of the maser locations depends on the quality of the 
phase calibration and is believed to be better than 0.5~arcsec. The absolute flux density scale
was bootstrapped from observation of Uranus (assumed flux densities were 1.38 and 1.98~Jy
at 36 and 44~GHz respectively).  We estimate the flux scale to be accurate to about 20\%. 
The correlator was configured to split 8~MHz bandwidth into 
1024~spectral channels.
The resulting spectral resolution and velocity range covered are given in Table~\ref{obsdetails}
along with other observing parameters. Most columns of Table~\ref{obsdetails} are self-explanatory
and include, in addition to above, the molecular transition and its rest frequency 
\citep[adopted from][]{mul04},
full width at half maximum (FWHM) and position angle (p.a.) of the synthesised beam reflecting the 
spatial resolution, as well as the size of the primary beam which determines the field of view. 
The 1$\sigma$ rms noise
given in the seventh column is a median value of rms noise levels obtained for each individual spectral plane
of the image cube prior to the primary beam correction (i.e. with constant noise across the field of view).
Scaled up with the primary beam, it represents the 1$\sigma$ detection limit of these observations at a particular location.
It is worth mentioning, however, that the ability to detect weak features is notably reduced within the 
velocity range of strong emission due to dynamic range limitations, which were largely a result of 
sparse uv-coverage. We estimate the dynamic range
(the ratio of peak flux to rms noise in the image) to be about 90 and 400 for the
36 and 44-GHz image cubes, respectively. 
The uncertainty of the radial velocity corresponding to the rest frequency uncertainty is given in the 
ninth column of Table~\ref{obsdetails}.  

\begin{table*}
 \caption{Observation details. The rms noise is calculated as a median of the noise levels obtained for each spectral plane of the image cube prior to the primary beam correction.
The velocity uncertainty corresponds to the uncertainty of the rest frequency, which is  given in the brackets following the value of the rest frequency and is expressed in the units of the least significant figure. }
 \label{obsdetails}
 \begin{tabular}{@{}lrclcccrcr}
 \cthead{Molecular}  & \cthead{UT date of} & \cthead{Rest} & 
 \multicolumn{2}{c}{Synthesised beam} &\cthead{Primary beam}& \cthead{$1\sigma$ rms} & \cthead{Velocity} & \cthead{Velocity} &\cthead{Spectral}\\ 
 \cthead{transition}&\cthead{observation}& \cthead{frequency} & \cthead{FWHM} &\cthead{p.a.} &
\cthead{FWHM} &\cthead{noise} &\cthead{range} & \cthead {uncertainty} &\cthead{resolution}\\
& & \cthead{(MHz)} &\cthead{(arcsec)}  & \cthead{(deg)}& \cthead{(arcmin)} & \cthead{(mJy)}&\cthead{(\ks)}&\cthead{(\ks)}&\cthead{(\ks)} \\
  \hline
$4_{-1}-3_0$~E & 23 May 2007& 36169.265~(30)  & 6.0$\times$4.7 & 76  & 1.40 & 100 & $-$85.4, $-$26.0 & 0.24\hphantom{0} & 0.066 \\
$7_{0}-6_1$~A$^+$  & 22 May 2007& 44069.410~(10) & 5.3$\times$4.0 & 80 & 1.15 &  130 & $-$76.9, $-$28.3 & 0.068 & 0.054\\
\hline
\end{tabular}
 \end{table*}

Data reduction in {\sc miriad} followed the standard procedure including an opacity 
correction based on the  model built into {\sc miriad} and atmospheric temperature, pressure and humidity registered at the 
time of observations by the ATCA weather station. We  used uniform weighting for the
imaging. After absolute positions had been determined we performed self-calibration (solution 
interval 1~min) using the brightest spectral feature as a reference. The flux density ratio of 
the reference  feature 
before and after self-calibration allowed us to estimate decorrelation caused by atmospheric 
phase variations occurring on a time-scale shorter than 5 minutes, the minimum temporal separation  
from the secondary calibrator scan. We estimated decorrelation factors to 
be 1.18$\pm$0.01 and 1.03$\pm$0.02 at 44 and 36~GHz, respectively. The full velocity range
was imaged using the self-calibrated dataset. Therefore,  to first order, the effect of the atmospheric phase stability on the derived flux densities had been corrected for.  The residual uncertainty is expected to be small compared to the accuracy of the flux scale calibration. The maser emission 
was searched in the cube prior to the primary beam correction (i.e. the noise across the field of view was constant). Then, the cube was divided by the primary beam model at the appropriate frequency and 
the spectra were extracted at the peak pixel by taking a slice along the spectral axis. We followed this
approach due to the rather high sidelobe level in the point spread function caused by the 
poor uv-coverage attained in the project. It reproduces flux density correctly in the case of unresolved or barely resolved sources  (i.e. smaller  than the synthesised beam) and is well suited to maser observations.

 \begin{figure*}
\includegraphics[width=\linewidth]{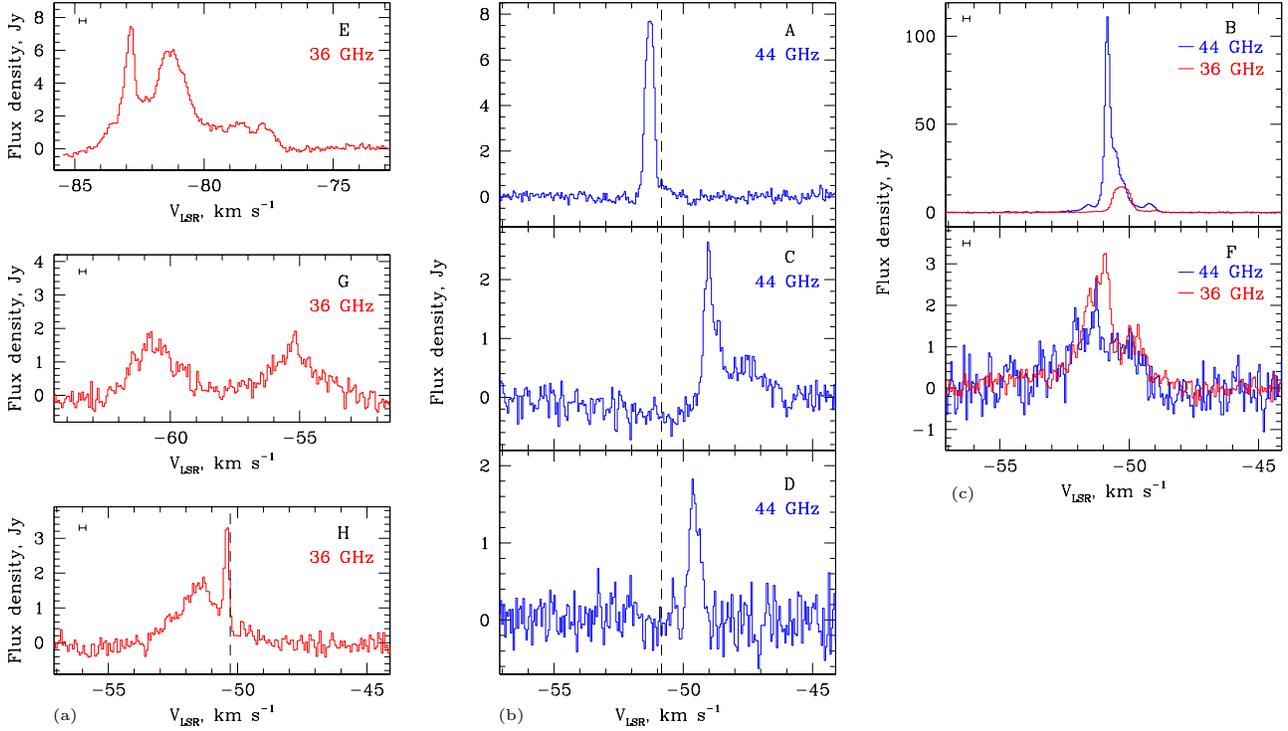}
\caption{Spectra of the 36 (red) and 44-GHz (blue) emission at eight locations in G309.38$-$0.13
(location~I is not shown here due to low signal-to-noise ratio in the spectrum). The dashed line
shows the peak velocity of the brightest spectral feature (at location B) at the given frequency.
(a)~Emission from E (high-velocity feature), G and H was detected at 36-GHz only. (b)~A, C and D were detected at 44~GHz only. 
(c)~B (main feature) and F were detected at both 36 and 44~GHz. The uncertainty of the relative velocity alignment between transitions is dominated by the rest frequency uncertainty of the 36-GHz transition shown by the horizontal error bar in spectra showing 36-GHz emission (it is 
comparable to the spectral resolution for the 44-GHz transition).}
\label{spectra}
\end{figure*}

\section{Results}
\label{results}

The morphology of the region is shown in Figure~\ref{glm_mas_overlay}.  The methanol emission
is confined to nine discrete locations, labelled A$-$I, spread 
over a 50\arcsec$\times$30\arcsec{ }region.
The spectra at most locations are complex, consisting of more than one spectral
feature (Fig.~\ref{spectra}). 
At finer scale (less than the synthesised beam size), each spectral feature corresponds to a 
slightly different position in the map. Most of the features are believed to be spots of class~I maser
emission for reasons summarised at the end of this section.
E~is at essentially the same position as~B, but
was designated by a separate letter as it corresponds to a distinct group of velocities, blue-shifted
by approximately 30~\ks. We show the positions of all spectral features corresponding to
each spot  (circles denote the brightest  feature, crosses show all other features) in 
Fig.~\ref{glm_mas_overlay} along with the position of the class~II methanol maser
at 6.7~GHz \citep{cas09}. The 6.7-GHz maser in this region peaks at $-$49.6~\kss and is 
quite weak (1~Jy).  It is found at $\alpha_{2000}=13^h47^m$23\fs98, $\delta_{2000}=-62$\degr18\arcmin12\farcs0, just 
a few seconds of arc south of the class~I maser emission at~B and~E. Both the OH and H$_2$O masers
are, within the measurement uncertainty, at the same position as the 6.7-GHz maser and have 
radial velocities near $-$50~\kss\citep{cas98,bre10}.
The 3-colour background in Fig.~\ref{glm_mas_overlay} shows the emission in the 
8.0, 4.5 and 3.6-$\mu$m IRAC Spitzer bands as red, green and blue, respectively. The infrared image reveals an EGO in the vicinity of~B and E, which is likely to be a signature of the shocked gas. A number of compact sources
showing excess in the 4.5-$\mu$m band (appearing green in Fig.~\ref{glm_mas_overlay}) 
are also present. In particular, such a source is located near~F and~G.  

The 
spectra at all locations except~I are shown in Figure~\ref{spectra}. The latter is not shown due to 
a poor signal to noise ratio, but qualified as a detection due to its
isolation in velocity from the other bright emission in the region, despite being comparable 
in flux density to dynamic range artefacts seen at other velocities. Only~B and F were reliably 
detected at both 
36 and 44~GHz.  E, G and H were detected at 36~GHz only, while only 44-GHz emission 
was detected at~A, C, D and~I.  To describe individual spectral features, 
we decomposed the spectra into a number of 
Gaussian components (the sum of the components represents each spectrum within measurement
errors). The LSR velocity, position, velocity FWHM of the component and the flux density
are given on the left-hand side of Table~\ref{fit_results}. The position fit was done assuming a point
source model, typically at the peak velocity for the appropriate component. In the case of 
significant blending, the position fit is expected to have a systematic 
error and give a flux density-weighted average position of the blended Gaussian components.  
We estimate that  the  systematic errors of the fitted position are likely to be less than the
random errors of the fit given in Table~\ref{fit_results} for most components in this particular source.
Two exceptions are the 30-Jy component towards~B at 44~GHz and the $-$78.6-\kss 
component  towards~E. Both were fitted with a single Gaussian
component, although the profile shape in Figure~\ref{spectra} suggests that at least two
independent components may be present in each case. 

The relatively compact ATCA configuration used for 
observations allows us to put only a relatively weak constraint on the brightness temperature 
($>$100~K),
which  is insufficient to prove the maser nature of the observed emission unambiguously. However,
given the line width of the majority of components (Table~\ref{fit_results} and Fig.~\ref{spectra}), at 
least some of the emission is likely to be due to maser action for most of the features. In addition,
we verified that the dominant emission at both~B (at both frequencies) and~E is unresolved
at baselines to distant CA06 antenna, which implies brightness temperatures exceeding 10$^6$~K,
and, therefore, a maser origin of this emission.
The right-hand side of Table~\ref{fit_results} shows 
the parameters of the line profile at each maser location, such as the LSR velocity and the
flux density of the peak as well as the integral over the line profile. 
 
\begin{table*}
\caption{Fit results and profile parameters. The uncertainties are
given in parentheses and expressed in units of the least significant
figure. Notes: ($a$) the uncertainty
is half of that for the line FWHM, ($b$) the uncertainty is the spectral
resolution listed in Table~\protect\ref{obsdetails}.}
\label{fit_results}
\begin{tabular}{@{}cl@{}c@{}llrrrlrr}
\hline
     & & & \multicolumn{5}{c}{Gaussian components} & \cthead{Peak} &
\cthead{Peak} \\
Spot &\cthead{Molecular}  & Frequency & \cthead{LSR} & \cthead{$\alpha_{2000}$} & \cthead{$\delta_{2000}$}&
\cthead{Line} & \cthead{Flux} &  \cthead{LSR} &
\cthead{flux} &\cthead{$\int f(v)\;dv$}\\
     & \cthead{transition} & & \cthead{Velocity\makebox[0mm]{\hskip 2mm $^a$}} & \cthead{13$^h$47$^m$} & \cthead{$-$62\degr} &
\cthead{FWHM} & \cthead{density} & 
\cthead{velocity\makebox[0mm]{\hskip 2mm $^b$}} &
\cthead{density} &\\
     & & (GHz) &\cthead{(\ks)} & \cthead{($^s$)}&\cthead{(arcmin~arcsec)} &
\cthead{(\ks)} & \cthead{(Jy)} &  \cthead{(\ks)} &
\cthead{(Jy)}&\cthead{(Jy \ks)}  \\
\hline

A & $7_0-6_1$~A$^+$ & 44 & 
$-$51.30\hphantom{10} &  24.97\hphantom{1}~(2) & 18~06.0\hphantom{11}~(1) &
0.387~(3) & 7.90\hphantom{6}~(6) &  $-$51.27 & 7.6~(2) & 3.21\hphantom{0}~(3)\\

B  & $4_{-1}-3_0$~E & 36 & $-$51.34\hphantom{0} &  24.0\hphantom{91}~(1) & 18~10.3\hphantom{11}~(6) &
0.42\hphantom{2}~(6) & 1.0\hphantom{6}~(1) &  $-$50.31 & 14.6~(1) &12.41\hphantom{0}~(3) \\
  & & & $-$50.295 &  24.141~(7) & 18~09.35\hphantom{1}~(4) &
0.715~(5) & 15.96~(9) &  \\
  & & & $-$49.06\hphantom{7} &  24.13\hphantom{1}~(3) & 18~08.6\hphantom{11}~(2) &
0.55\hphantom{2}~(6) & 1.1\hphantom{6}~(1) &  \\

  & $7_0-6_1$~A$^+$  & 44 & $-$51.6\hphantom{10} &  24.13\hphantom{1}~(3) & 18~09.3\hphantom{11}~(2) &
0.5\hphantom{29}~(3) & 3.5\hphantom{6}~(2) & $-$50.84 & 111.1~(2) & 50.82\hphantom{0}~(6)\\
  & & & $-$50.851 &  24.095~(2) & 18~09.165~(8) &
0.192~(1) & 85.9~(3) &  \\
  & & & $-$50.58\hphantom{0} &  24.110~(4) & 18~09.02\hphantom{1}~(2) &
0.78\hphantom{2}~(3) & 30.7~(7) &  \\
  & & & $-$49.23\hphantom{0} &  24.194~(8) & 18~08.56\hphantom{1}~(4) &
0.6\hphantom{22}~(2) & 4.4\hphantom{6}~(9) &  \\

C  & $7_0-6_1$~A$^+$ & 44 & $-$49.01\hphantom{0} &  24.05\hphantom{1}~(2) & 18~01.16\hphantom{1}~(9) &
0.51\hphantom{0}~(3) & 1.95\hphantom{3}~(7) &  $-$49.04 & 2.6~(2) & 2.12\hphantom{0}~(9)\\
  & & & $-$47.83\hphantom{0} &  24.09\hphantom{1}~(9) & 18~01.6\hphantom{10}~(4) &
1.7\hphantom{10}~(2) & 0.49\hphantom{3}~(4) &  \\

D & $7_0-6_1$~A$^+$ & 44 & $-$49.57\hphantom{0} &  22.88\hphantom{4}~(2) & 17~51.4\hphantom{11}~(2) &
0.55\hphantom{0}~(3) & 1.54\hphantom{3}~(7) & $-$49.64 & 1.8~(3) & 0.81\hphantom{0}~(2)\\

E  &$4_{-1}-3_0$~E & 36 & $-$83.50\hphantom{1} &  24.14\hphantom{1}~(3) & 18~10.4\hphantom{31}~(2) &
0.75\hphantom{9}~(5) & 1.49\hphantom{3}~(9) &  $-$82.84 & 7.5~(2) & 18.27\hphantom{0}~(8) \\
  & & & $-$82.854 &  24.085~(8) & 18~10.47\hphantom{1}~(5) &
0.485~(9) & 6.5\hphantom{36}~(1) &  \\
  & & & $-$81.33\hphantom{1} &  24.080~(3) & 18~10.52\hphantom{1}~(2) &
1.64\hphantom{1}~(2) & 6.5\hphantom{36}~(1) &  \\
  & & & $-$78.60\hphantom{1} &  24.06\hphantom{1}~(3) & 18~10.4\hphantom{31}~(2) &
2.28\hphantom{9}~(9) & 1.44\hphantom{3}~(5) &  \\

F  & $4_{-1}-3_0$~E & 36 & $-$52.0\hphantom{15} &  20.64\hphantom{1}~(8) & 18~05.7\hphantom{11}~(5) &
5.0\hphantom{92}~(3) & 0.51\hphantom{6}~(2) & $-$50.92 & 3.3~(2) &5.49\hphantom{0}~(2) \\
  & & & $-$51.07\hphantom{5} &  20.64\hphantom{1}~(5) & 18~04.6\hphantom{11}~(3) &
1.33\hphantom{2}~(2) & 2.66\hphantom{6}~(4) &  \\
  & & & $-$49.66\hphantom{5} &  20.75\hphantom{1}~(4) & 18~04.3\hphantom{11}~(3) &
0.67\hphantom{2}~(4) & 1.12\hphantom{6}~(5) &  \\

  & $7_0-6_1$~A$^+$ & 44 & $-$51.76\hphantom{1} &  20.5\hphantom{21}~(2) & 18~06.8\hphantom{11}~(7) &
1.35\hphantom{2}~(9) & 1.26\hphantom{6}~(7) & $-$51.27 & 2.7~(4) & 3.5\hphantom{10}~(2)\\
  & & & $-$49.62\hphantom{1} &  20.7\hphantom{21}~(2) & 18~06\hphantom{.111}~(1) &
1.1\hphantom{12}~(2) & 0.57\hphantom{6}~(8) &  \\

G & $4_{-1}-3_0$~E  & 36 & $-$60.56\hphantom{1} &  19.27\hphantom{1}~(5) & 18~11.5\hphantom{11}~(3) &
1.98\hphantom{1}~(8) & 1.54\hphantom{6}~(6) & $-$55.16 & 1.9~(4) & 7\hphantom{.100}~(1) \\
  & & & $-$55.29\hphantom{1} &  19.32\hphantom{1}~(4) & 18~11.6\hphantom{11}~(3) &
2.2\hphantom{19}~(1) & 1.31\hphantom{6}~(5) &  \\

H  & $4_{-1}-3_0$~E & 36 & $-$51.50\hphantom{1} &  21.24\hphantom{1}~(3) & 17~55.0\hphantom{11}~(2) &
2.04\hphantom{1}~(5) & 1.48\hphantom{6}~(3) & $-$50.39 & 3.3~(2) & 3.79\hphantom{0}~(5) \\
  & & & $-$51.432 &  21.52\hphantom{1}~(2) & 17~56.9\hphantom{11}~(1) &
0.275~(8) & 3.30\hphantom{6}~(9) &  \\

I & $7_0-6_1$~A$^+$ & 44 & $-$53.25\hphantom{9} &  25.19\hphantom{1}~(5) & 18~10.8\hphantom{11}~(2) &
0.81\hphantom{1}~(8) & 0.43\hphantom{6}~(4) & $-$53.29 & 0.6~(2) & 0.404~(8)\\
\hline
\end{tabular}
\end{table*}

\section{Discussion}
\subsection{High-velocity maser feature}
\label{hv_discussion}

The blue-shifted spectral component corresponding to location~E is a remarkable feature of this
maser source. More commonly, the observed
velocities of class~I methanol masers are no more than a few \kss from the velocity of the ambient 
molecular cloud, and the velocity dispersion is low~\citep[e.g.,][]{bac90,pla90,vor06,ara09}. The 
notable exception with a high velocity dispersion is G1.6$-$0.025, which has emission components
at radial velocities ranging from 0 to 160~\ks~\citep{has93,sal02}. However, unlike G309.38$-$0.13,
the emission in G1.6$-$0.025 corresponds to distinct, but interacting, molecular clumps 
detected also in a number of thermal transitions of various molecules~\citep{sal02}. The large
velocity difference between individual clumps reflects a violent environment and the exceptional
kinematics in the vicinity of the Galactic centre. 

In contrast,  the G309.38$-$0.13 region
is close to the tangential direction of the Sagittarius-Carina arm \citep[see Figure~5 of][]{rus03}
where the extreme velocity of~E ($-$80~\ks) makes an association with an unrelated molecular
cloud highly unlikely.
This is corroborated by the $^{12}$CO brightness temperature limit of 0.1~K at the radial velocities
below $-$70~\kss obtained with the 4-m Nanten telescope~\citep{sai01}. This limit is two orders
of magnitude lower than the peak $^{12}$CO brightness temperature observed near 
$-$50~\kss in the G309.38$-$0.13 region \citep[nearest $^{18}$CO clump mass is about 
10$^4$~M$_\odot$;][]{sai01}, which is close to the velocities of most of the maser spots.

We conclude that  the high-velocity 36-GHz maser in~E most likely reflects the presence of 
a molecular outflow, rather than a separate molecular cloud projecting onto the same location
by chance. 
The area around~B and~E is characterised by the presence of shocked gas traced by a strong EGO 
(Fig.~\ref{glm_mas_overlay}). This supports the idea that an outflow is present. The close spatial
location of~B and~E and the presence of the class~II methanol maser 
(which indicates the presence of a high-mass YSO) in the vicinity both suggest that the outflow axis is 
close to the line of sight. 

As mentioned earlier, the interaction between outflows and the ambient material is believed to be
responsible for  the majority of class~I methanol masers.  The production of methanol by purely 
gas-phase chemical reactions is quite inefficient, but the passage of weak shocks 
is expected to release methanol from the grain mantles \citep[e.g.,][]{har95}. 
Enhanced methanol abundance arising from shocks driven by
outflows into the ambient cloud has also been confirmed observationally~\citep[e.g.,][]{gib98,gar02}. 
In addition, outflows compress the medium, which makes collisions more frequent and the pumping mechanism more efficient. However, methanol is expected to survive sputtering and desorption of grain
mantles only if the shocks are relatively mild, i.e. at shock velocities not greatly exceeding 
10~\ks~\citep{gar02}. Therefore, the detection of a maser feature offset by 30~\kss with respect to the
velocity of the ambient gas is somewhat surprising. 

This discrepancy can be resolved if the outflow interacts with a moving parcel of gas.
In this case, the same radial velocity of the maser can be achieved for a lower shock velocity.
The mechanism of this gas acceleration is unknown at present. It could be by a previous outflow 
event or a different outflow from another YSO.  In addition, the clustered environment of
high-mass star formation opens a range of possibilities for gravitational acceleration of gas 
clumps~\citep[e.g. mechanisms which caused radial velocity outliers observed in clusters of low-mass 
protostars by][]{cov06}. 
One may speculate that
the curved morphology of the EGO associated with~B and~E (Fig.~\ref{glm_mas_overlay})
suggests interaction with an outflow coming from the south-west (the general direction
of the H{\sc ii} region Gum~48d). This may be the 
additional outflow responsible for the unusual kinematics. However, the distance of Gum~48d 
is estimated to be 3.5~kpc \citep{kar09}, and therefore Gum~48d must be at least 
20~pc away from the maser. A separation of the terminal shock 
from the driving source as large as 
this is unprecedented (e.g. the outflow studied by \citet{bro03} extends only up to 1.5~pc) 
and suggests that, if the 
EGO is indeed shaped by the second outflow, the driving source is not the star responsible for the excitation of Gum~48d.
 
\subsection{Relative flux density of 36 and 44-GHz methanol masers}
An inspection of Table~\ref{fit_results} and Fig.~\ref{spectra} suggests that the 36 to 44-GHz 
flux density ratio varies significantly from component to component. The majority of maser features 
were detected in a single transition only and most striking is the lack of a 36-GHz counterpart
to the brightest 44-GHz spectral feature (Fig~\ref{spectra}c). Although the velocities
corresponding to~E (high-velocity feature) were not covered by the 44-GHz measurement, the
preliminary results of a 
follow up observation carried out with the Mopra 22-m single dish in January 2010 showed
no 44-GHz methanol (as well as no 48-GHz methanol, SiO 1-0, and CS 1-0) emission 
near $-$80~\kss above 1.3~Jy. With the caveat about possible temporal variability
over the 2.5-year timescale, this situation represents another extreme case of a 36-GHz maser without
a 44-GHz counterpart. The numerical models of \citet{sob05} suggest that the 44-GHz 
masers are quite sensitive to the geometry of the maser region. Unlike 36-GHz masers, which
can be produced in the spherically symmetric case, an optical depth along the line of sight higher 
than in the orthogonal direction seems to be the key property required to produce bright 44-GHz 
masers. One could expect this elongation of the gas clumps involved in the maser action to be 
somehow related to the outflow orientation and morphology. However, the complexity of the maser
spot distribution (see Fig.~\ref{glm_mas_overlay}) does not allow us to interpret the flux density
ratios for each spot in a consistent manner based on just the spot location with respect to the 
suspected outflow (the position of the 6.7-GHz maser presumably marks the location of the YSO 
driving the outflow roughly along the line of sight).

\section{Conclusions}
\begin{enumerate}
\item Spots of class~I methanol maser emission in G309.38$-$0.13  are
spread over an area 50\arcsec$\times$30\arcsec{ }in extent (see Table~\ref{fit_results} and 
Fig.~\ref{glm_mas_overlay}). Both 36 and 44-GHz maser emission was detected at two
locations, while the emission from the remaining locations was detected at a single 
frequency only (see Fig.~\ref{spectra}).

\item We report the detection of a high-velocity spectral feature at 36~GHz, blue-shifted by about 30~\kss
from the peak velocity. This is the largest velocity offset reported so far for a class~I methanol 
maser source 
associated with a single molecular cloud. The maser corresponding to this feature (labelled~E) 
is located within a few seconds of arc of the maser at~B (the brightest spectral 
feature at both frequencies) indicating a possible association with an outflow 
parallel to the line of sight.

\item The results highlight the importance of not only high spectral and spatial resolution 
for studies of class~I methanol masers, but also ample velocity coverage and a field of 
view wider than the typical primary beam size of available instruments.  All observational data 
gathered so far lack at least one (and sometimes all) of  these properties. In addition, the results
show that the 36 and 44-GHz transitions are very complementary to each other.
The high-velocity spectral feature in G309.38$-$0.13 is a good example of a phenomenon 
which has been missed by previous maser surveys. 

\end{enumerate}

\section*{Acknowledgments}
The Australia Telescope is funded by the
Commonwealth of Australia for operation as a National Facility managed
by CSIRO. SPE thanks the Alexander-von-Humboldt-Stiftung for an Experienced Researcher Fellowship which has helped support this research.  
AMS was financially supported by the Russian Foundation for Basic
Research (grant 10-02-00589-a) and the Russian federal program
``Scientific and scientific-pedagogical  personnel of innovational
Russia'' (contracts 02.740.11.0247 from 07.07.2009 and 540 from
05.08.2009). The research has made use of the NASA/IPAC Infrared
Science Archive, which is operated by the Jet Propulsion Laboratory,
California Institute of Technology, under contract with the National
Aeronautics and Space Administration.

\bsp

\label{lastpage}

\end{document}